\documentclass[aps,prd,noshowpacs,nofootinbib,noshowkeys,floatfix,superscriptaddress,singlecolum]{revtex4}
\usepackage[dvips]{graphics,graphicx}
\usepackage[colorlinks=true,linktocpage=true,linkcolor=blue,citecolor=blue]{hyperref}
\usepackage[usenames,dvipsnames]{color}
\usepackage{amsmath, amssymb,oldgerm}
\usepackage{multirow}
\usepackage{longtable}
\usepackage{color}
\usepackage[normalem]{ulem}  
\usepackage{braket}
\usepackage{slashed}
\usepackage[mathscr]{eucal}

\newcommand{\beq}{\begin{equation}}
\newcommand{\eeq}{\end{equation}}

 \newcommand{\bbm}{\begin{bmatrix}}
\newcommand{\ebm}{\end{bmatrix}}
\newcommand{\MA}{\mathcal{A}}

\newcommand{\bea}{\begin{eqnarray}}
\newcommand{\eea}{\end{eqnarray}}

\begin{document}

\title{Challenges in Solving Chiral Hydrodynamics}

\author{Enrico Speranza}
\email{espera@illinois.edu}
\affiliation{Illinois Center for Advanced Studies of the Universe \& Department of Physics, 
University of Illinois at Urbana-Champaign, Urbana, IL, 61801, USA}

\author{F\'abio S.\ Bemfica}
\email{fabio.bemfica@ect.ufrn.br}
\affiliation{Escola de Ci\^encias e Tecnologia, Universidade Federal do Rio Grande do Norte, RN, 59072-970, Natal, Brazil}

\author{Marcelo M.\ Disconzi}
\email{marcelo.disconzi@vanderbilt.edu}
\affiliation{Department of Mathematics, Vanderbilt University, Nashville, TN, 37211, USA}

\author{Jorge Noronha}
\email{jn0508@illinois.edu}
\affiliation{Illinois Center for Advanced Studies of the Universe \& Department of Physics, 
University of Illinois at Urbana-Champaign, Urbana, IL, 61801, USA}

\begin{abstract}
We prove that ideal chiral hydrodynamics, as derived from chiral kinetic theory, is acausal and its initial-value problem is ill-posed both in the linearized case around a local equilibrium solution and also in the full nonlinear regime. Therefore, such theory cannot be used to determine how the chiral anomaly affects the hydrodynamic evolution. We show that these fundamental issues can be fixed by using different definitions (frames) for the hydrodynamic fields. This leads to a causal theory of ideal chiral hydrodynamics where the vorticity strength is constrained by the coefficient that encodes the anomaly.
\end{abstract}

\maketitle

\noindent
\section{Introduction} Relativistic fluid dynamics  \cite{Rezzolla_Zanotti_book} is a fundamental tool in various fields ranging from high-energy nuclear physics \cite{Heinz:2013th,Romatschke:2017ejr,Florkowski:2017olj} to astrophysics  \cite{Baiotti:2016qnr}. Considerable effort has been made to study the novel phenomena displayed by chiral systems arising from the interplay between quantum  anomalies and the electromagnetic and vortical fields. Such effects can influence the dynamics of various systems, from the quark-gluon plasma to Weyl semimetals \cite{Kharzeev:2015znc,Huang:2015oca,Hosur:2013kxa}. While the chiral magnetic effect \cite{Kharzeev:2007jp,Fukushima:2008xe} is explicitly connected to the axial anomaly, the microscopic origin of the chiral vortical effect \cite{vilenkin1979macroscopic,vilenkin1980quantum,Erdmenger:2008rm,Banerjee:2008th,Son:2009tf,Landsteiner:2011cp,Landsteiner:2011iq} may be understood in different ways \cite{Glorioso:2017lcn,Flachi:2017vlp,Avkhadiev:2017fxj,Stone:2018zel,Buzzegoli:2018wpy,Prokhorov:2020okl,Huang:2020kik}.

The chiral vortical effect was first investigated in \cite{vilenkin1979macroscopic,vilenkin1980quantum} and later in  relativistic hydrodynamics in Refs. \cite{Erdmenger:2008rm,Banerjee:2008th} in the context of the fluid/gravity duality \cite{Bhattacharyya:2008jc}, which predicted the inclusion of a new term in the standard constitutive relations for the dissipative currents. In Refs.\ \cite{Son:2009tf,Sadofyev:2010pr,Neiman:2010zi}  an entropy-current analysis was used to reach a similar conclusion showing that the presence of quantum anomalies modifies the hydrodynamic equations, and the associated transport phenomena can occur even without dissipation. Therefore, in chiral (or anomalous) relativistic hydrodynamics, subtle quantum effects driven by anomalies can become manifest even in the macroscopic regime. 

It is possible to derive the equations of motion of chiral hydrodynamics from a kinetic theory formulation \cite{Chen:2015gta}, the so-called chiral kinetic theory (see also \cite{Son:2012wh,Stephanov:2012ki,Chen:2012ca,Manuel:2013zaa,Chen:2014cla,Manuel:2014dza,Gorbar:2017cwv}) which, in turn, can be obtained from quantum field theory using the Wigner function \cite{Son:2012zy,Hidaka:2016yjf,Hidaka:2017auj,Huang:2018wdl,Gao:2018wmr,Yang:2018lew,Gao:2018jsi,Carignano:2018gqt} and the world-line formalism \cite{Mueller:2017lzw,Mueller:2017arw}. While these approaches have provided great insight into the physics of chiral  matter, very little is known about the properties of the chiral hydrodynamic equations of motion  and their solutions, especially in the nonlinear regime. Such knowledge is relevant when studying the consequences of the chiral anomaly in  hydrodynamic simulations of the quark-gluon plasma formed in heavy-ion collisions \cite{Shi:2017cpu,Shi:2019wzi}.

In this work we take essential steps towards solving this issue by investigating the initial-value problem of ideal (i.e., dissipationless) chiral hydrodynamics, as derived from chiral kinetic theory \cite{Chen:2015gta}.
We prove that the initial-value problem for these equations of motion is ill-posed, both in the linear and nonlinear regimes. In other words, given arbitrary initial data there does not exist a corresponding solution to the equations of motion or a solution is not unique. Also, causality is violated in such a theory. Therefore, our analysis implies that it is hopeless to implement such equations of motion in numerical codes that simulate the hydrodynamic evolution of the quark-gluon plasma. We show that this issue can be fixed by considering different definitions for the hydrodynamic fields (i.e., different hydrodynamic frames), following  \cite{Bemfica:2017wps,Kovtun:2019hdm,Bemfica:2019knx,Hoult:2020eho,Bemfica:2020zjp}. This procedure leads to a causal formulation of ideal chiral hydrodynamics where the magnitude of the vorticity is constrained by the quantum anomaly coefficient.
For completeness, we also study the standard formulation of viscous chiral hydrodynamics defined at first-order in gradients \cite{Son:2009tf} and show that this theory also violates causality.

This paper is organized as follows. In Section \ref{secII} we present the equations of motion of ideal chiral hydrodynamics as derived from kinetic theory. Section \ref{sec:initial} we study the initial value problem of this theory in the linearized regime (around a solution of ideal hydrodynamics) and also in the full nonlinear regime. We carefully examine both the linear and nonlinear cases in this paper in a comprehensive manner, so that the reader can clearly understand the regime of validity of the statements made in either case.  In Section \ref{secIV} we perform a similar study in the case where the equations are written in the Landau hydrodynamic frame. We investigate causality in the nonlinear regime of standard chiral viscous hydrodynamics in Section \ref{secV}. Our conclusions and outlook are presented in Section \ref{conclusions}, and in Appendix \ref{appendixA} we list the linearized equations of motion used in Section \ref{sec:initial}. \emph{Notation}: We use a mostly plus Minkowski metric $g_{\mu\nu} = \mathrm{diag}(-,+,+,+)$, natural units $\hbar=k_B=c=1$, and Greek indices run from 0 to 3. 

\section{Equations of motion of ideal chiral hydrodynamics from kinetic theory} 
\label{secII}

We first consider chiral hydrodynamics in the dissipationless regime, which can be directly derived using chiral kinetic theory \cite{Chen:2015gta}. The starting point  involves defining the distribution function that describes local equilibrium. If the collisional invariants are energy, momentum, charge and total angular momentum, the equilibrium distribution function for massless fermions takes the form (the extension to antifermions is straightforward) \cite{Chen:2015gta}
$
f_{\text{eq,}\pm}(x,p)=[{\exp ({g_\pm})+1}]^{-1},
$
where 
$
g_\pm (x,p)=- u^\mu  p_\mu/T -  {\mu_\pm}/T \mp \frac{1}{4} S^{\mu\nu}\varpi_{\mu\nu},
$
with $u^\mu$ being the relativistic flow velocity (with $u_\mu u^\mu = -1$), $T$ the temperature, $\mu_\pm$ the chemical potential for right and left-handed particles, $\varpi^{\mu\nu}=-(1/2)[\partial_{\mu}(u_\nu/T)-\partial_{\nu}(u_\mu/T)]$ the thermal vorticity, and $S^{\mu\nu}= \epsilon^{\mu\nu\rho\sigma}p_\rho n_\sigma/(p_\alpha n^\alpha)$
the dipole-moment tensor which encodes the Lorentz frame dependence through the frame vector $n^\mu$ related to the side-jump effect \cite{Chen:2014cla,Chen:2015gta,Stone:2014fja,Stone:2015kla}. Also, $p_\mu$ is the particle 4-momentum. Using the distribution function in phase space given above, one can employ an $\hbar$-gradient expansion to obtain the constitutive relations that define the macroscopic quantities. At first-order, one obtains \cite{Chen:2015gta}
\begin{subequations}
\label{css}
\begin{align}
T^{\mu\nu}&= \varepsilon u^\mu u^\nu + P\Delta^{\mu\nu} + \xi_T (\omega^\mu u^\nu + \omega^\nu u^\mu), \\
J^\mu_V&= n_V u^\mu + \xi_V\omega^\mu, \\
J^\mu_A&= n_A u^\mu + \xi_A\omega^\mu,
\end{align}
\end{subequations}
where $T^{\mu\nu}$ is the energy-momentum tensor, $J^\mu_V$ is the vector current, $J^\mu_A$ is the axial-vector current, $\varepsilon = u_\mu u_\nu T^{\mu\nu}$ is the energy density, $P=\varepsilon/3$ is the equilibrium pressure, and $n_{V/A}$ are the vector (V) and axial-vector (A) densities, respectively. We also introduced above $\Delta_{\mu\nu} = g_{\mu\nu}+u_\mu u_\nu$, which is a tensor projector orthogonal to the flow, and the vorticity tensor
\begin{equation}
\omega^{\mu}=\frac{1}{2}\epsilon^{\mu\nu\alpha\beta}u_\nu \partial_\alpha u_\beta,
\label{vorticitydef}
\end{equation} 
where $\epsilon^{\mu\nu\alpha\beta}$ is the Levi-Civita symbol.

The essential novelty that distinguishes this chiral theory from ideal relativistic hydrodynamics \cite{Rezzolla_Zanotti_book} is the explicit presence of the vorticity tensor $\omega^\mu$ in the definition of the quantities in \eqref{css}, accompanied by the coefficients $\xi_T$, $\xi_V$, and $\xi_A$ (which vanish for nonanomalous matter). Thus, even though there is no dissipation in this theory, a nonzero energy flux given by $-\Delta^\mu_\lambda T^{\lambda\nu}u_\nu = \xi_T \omega^\mu$ is present, and the currents possess contributions transverse to the flow velocity (note that $\omega_\mu u^\mu=0$). These new contributions, which stem from the quantum anomaly, have important physical and mathematical consequences to the evolution of the fluid, as we discuss below.
The coefficients $\xi_T$, $\xi_V$, and $\xi_A$ are first-order in the $\hbar$-gradient expansion and they represent quantum corrections to the motion of the fluid. Furthermore, the terms with coefficients $\xi_V$ and $\xi_A$ constitute the chiral vortical and the axial-chiral vortical effects, respectively. The coefficients in \eqref{css} can be explicitly computed from the equilibrium distribution function, see e.g., Refs.\ \cite{Chen:2015gta,Yang:2018lew}.

The hydrodynamic equations of motion associated with the constitutive relations in \eqref{css} are the conservation laws, i.e., energy-momentum conservation $\partial_\mu T^{\mu\nu}=0$, and the conservation of the currents, $\partial_\mu J^{\mu}_V=0$ and $\partial_\mu J^{\mu}_A=0$. Without any loss of generality, in our analysis we use the projections of  $\partial_\mu T^{\mu\nu}=0$ parallel and orthogonal to $u^\mu$, i.e., we write the set of equations of motion as
\begin{subequations}
\label{csseoms}
\begin{align}
 u_\nu\partial_\mu T^{\mu\nu}= - D \varepsilon - (\varepsilon +P)\theta - \partial_\mu (\xi_T \omega^\mu) + u_\mu D (\xi_T \omega^\mu)&=0,  \\
 \Delta^\alpha_\nu\partial_\mu T^{\mu\nu}= (\varepsilon + P) D u^\alpha + \Delta^{\alpha\mu}\partial_\mu P + \xi_T \omega^\nu \partial_\nu u^\alpha + \Delta^\alpha_\nu D (\xi_T \omega^\nu) + \xi_T \omega^\alpha \theta &=0, \\
 \partial_\mu J^\mu_V = D n_V +n_V \theta + \xi_V\partial_\mu \omega^\mu +\omega^\mu\partial_\mu \xi_V &=0, \\
 \partial_\mu J^\mu_A = D n_A +n_A \theta + \xi_A\partial_\mu \omega^\mu +\omega^\mu\partial_\mu \xi_A  &=0 ,
\end{align}
\end{subequations}
where $D=u_\mu \partial^\mu$ and $\theta=\partial_\mu u^\mu$.
Below, we study the initial-value problem of Eqs.~\eqref{csseoms} in the linear and fully nonlinear regimes.

\section{Initial-value problem of ideal chiral hydrodynamics}
\label{sec:initial}

We prove below that the initial-value problem of the system of partial differential equations (PDEs) in  \eqref{csseoms} is ill-posed and the dynamics is necessarily acausal. 
The proof is based on standard techniques from the theory of PDEs \cite{Courant_and_Hilbert_book_2}. Consider a general quasilinear system of PDEs,
\begin{equation}
\label{gen_pde}
\mathcal{A}(\Psi,\partial)\Psi=\mathcal{B},
\end{equation}
where $\Psi(x) \in\mathbb{R}^N$ is a column vector of the $N$ unknowns of the system, $\mathcal{A}(\Psi,\partial)$ defines the so-called principal part of the system of PDEs \cite{ChoquetBruhatGRBook} which corresponds to an $N\times N$ matrix differential operator, possibly depending on $\Psi$ and its derivatives, containing the higher-order derivatives of each unknown in the system (the order of the higher derivative of different unknowns need not be the same), and $\mathcal{B}$ is a column vector that may depend on the unknowns as well as their lower-order derivatives. The initial-value problem consists of finding a solution 
of the system  \eqref{gen_pde} with given initial values of $\Psi$ and their lower-order derivatives along a hypersurface $\Sigma$, which one can parametrize as $\phi(x)=0$ (in most physical contexts, it is convenient to choose the hypersurface of vanishing initial time, i.e., $\phi(x)=x^0$). The initial-value problem is locally well-posed 
if for arbitrary initial data on $\Sigma$ there exists a unique solution\footnote{One often also requires that solutions vary continuously with the initial data, but here we focus only on
existence and uniqueness as these are essential features for physical theories.} in a neighborhood of $\Sigma$. 
For relativistic theories causality must also hold \cite{ChoquetBruhatGRBook}. Examples of theories of relativistic fluid dynamics where the initial-value formulation has been proven to be locally well-posed and strongly hyperbolic \cite{Reula:2004xd}, in the full nonlinear regime, are the ideal relativistic fluid \cite{ChoquetBruhatGRBook}, Israel-Stewart theory \cite{MIS-6} including only bulk viscosity effects \cite{Bemfica:2019cop}, and the generalized first-order theories of viscous hydrodynamics \cite{Bemfica:2020zjp}.

A minimal requirement for well-posedness is that one should be able to express the higher-order derivatives in terms of the lower-order derivatives, so that one can recursively determine all derivatives of a solution in terms
of the initial data. This will not be possible if for any covector $\varphi_\mu = \partial_\mu \phi$, where $\Sigma = \{
\phi(x) = 0 \}$, the characteristic determinant vanishes, i.e.,
\begin{equation}
\label{car}
\det [\mathcal{A}(\Psi_0,\varphi)] =0,
\end{equation}
where $\Psi_0$ is the initial data (i.e., the values of $\Psi$ and its lower-order derivatives along $\Sigma$).
In this situation the initial-value problem is locally ill-posed, i.e., given arbitrary $\Psi_0$, 
a solution of  \eqref{gen_pde} either does not exist or, if it does, it is not unique \cite{Hadamard_Book}.
This is a strongly undesirable feature to be displayed by fluid dynamic theories,  which are supposed to give 
unique physical solutions that lead to testable predictions.

Causality is verified by means of the system's characteristics, which are the roots of $\det[\MA(\Psi,\varphi)]=0$, where the replacement $\partial	_\mu\to \varphi_\mu$ has to be applied. The system is causal when the roots $\varphi_\mu=(\varphi_0(\varphi_i),\varphi_i)$ are such that \cite{Courant_and_Hilbert_book_2}
\begin{equation}
\label{condvarphi}
    \text{(i)}\,\, \text{$\varphi_\mu$ is real}  \qquad \text{and} \qquad \text{(ii)}\,\, \varphi^\mu\varphi_\mu\ge 0.
\end{equation}
For the sake of illustration, we give here two basic examples that will be relevant for the discussion below. 
\begin{itemize}
    \item Let us consider the advection equation
    \begin{equation}
    \label{advection}
        V_\mu \partial^\mu \Psi =0 ,
    \end{equation}
    where $\Psi$ is here one scalar function and $V_\mu$ is a general vector. The principal part is thus given by $\mathcal{A}(\Psi,\partial)=V_\mu\partial^\mu$. Following the prescription given above, we need to study the roots of $\det[\MA(\Psi,\varphi)]=V_\mu \varphi^\mu=0$. One can check that the conditions in \eqref{condvarphi} are satisfied and, thus, the system is causal, if and only if $V^\mu$ is timelike or lightlike, namely $V^\mu V_\mu \le 0$.

    \item Let us now consider the wave equation 
    \begin{equation}
    \label{wave}
        u_\mu u_\nu \partial^\mu \partial^\nu \Psi - \beta \Delta_{\mu\nu}\partial^\mu \partial^\nu \Psi=0,
        \end{equation}
        where again $\Psi$ is a scalar unknown and $\beta$ is a coefficient. The principal part is thus given by $\mathcal{A}(\Psi,\partial)=(u_\mu u_\nu   - \beta \Delta_{\mu\nu})\partial^\mu \partial^\nu$. As shown in \cite{Bemfica:2020xym}, any root $\varphi_\mu$ from the equation $\det[\MA(\Psi,\varphi)]=(u_\mu\varphi^\mu)^2   - \beta \Delta_{\mu\nu}\varphi^\mu\varphi^\nu=0$ obeys (i) and (ii) if, and only if, $0\le\beta\le1$. This follows from $\varphi^\mu \varphi_\mu=-(u_\mu\varphi^\mu)^2   +  \Delta_{\mu\nu}\varphi^\mu\varphi^\nu=(1-\beta)\Delta_{\mu\nu}\varphi^\mu\varphi^\nu$.
\end{itemize}

We shall now use the concepts discussed above to prove that the initial-value problem of  ideal chiral hydrodynamics \eqref{csseoms} is ill-posed in the sense of \eqref{car}. 
The first step consists in recognizing that the nonlinear set of second-order PDEs in \eqref{csseoms} is of the form  \eqref{gen_pde}.

\subsection{Linear regime}
\label{LinearChiral}

Let us first study the initial-value problem of the system \eqref{csseoms} in the linearized regime. We consider the linearization around a general (nonlinear) solution of conventional ideal hydrodynamics. We stress that the background solution here does not need to be the one describing the global equilibrium state.  Perturbation around such a state can be expressed in the following form,
\begin{equation}
\label{pert}
    \varepsilon=\varepsilon^{(0)}+\delta\varepsilon, \quad u^\mu = u^{(0)\mu}+\delta u^\mu, \quad \omega^\mu = \omega^{(0)\mu}+\delta\omega^\mu, \quad   n_V=n^{(0)}_{V/A}+\delta n_{V/A}, \quad \xi_{V/A}=\xi_{V/A}^{(0)}+\delta \xi_{V/A},
\end{equation}
where $\omega^{(0)\mu}=(1/2)\epsilon^{\mu\nu\alpha\beta}u^{(0)}_\nu\partial_\alpha u^{(0)}_\beta$ and 
\begin{equation}
    \delta \omega^\mu =\frac12 \epsilon^{\mu\nu\alpha\beta} ( u^{(0)}_\nu \partial_\alpha \delta u_\beta + \delta u_\nu \partial_\alpha u^{(0)}_\beta).
\end{equation}
The background fields $\varepsilon^{(0)}, \, u^{(0)\mu}, \, \omega^{(0)\mu}, \, n^{(0)}_{V/A},$ and  $\xi^{(0)}_{V/A}$ are, in general, spacetime dependent and satisfy the equations of motion of ideal hydrodynamics
\begin{align}
    D^{(0)} \varepsilon^{(0)} + \frac43 \varepsilon^{(0)}\partial_\alpha u^{(0)\alpha} &=0, \\
 4\varepsilon^{(0)} D^{(0)} u^{(0)\mu} + \Delta^{(0)\mu\alpha}\partial_\alpha \varepsilon^{(0)} &=0, 
\end{align}
where we used the conformal equation of state and defined $D^{(0)}=u^{(0)}_\alpha\partial^\alpha$ and $\Delta^{(0)\mu\nu}=g^{\mu\nu}+u^{(0)\mu}u^{(0)\nu}$. Note that the background fields that we use include the global equilibrium state as a particular case. 
Plugging Eq.~\eqref{pert} into Eqs.~\eqref{csseoms} and keeping only terms at first order in the perturbations, we obtain the linearized system of chiral hydrodynamics. The explicit form of such equations is quite lengthy and is given in App. \ref{appendixA}. One can show that such system 
can be cast in the form of \eqref{gen_pde} with the unknowns $\Psi=(\delta\varepsilon, \delta n_V, \delta n_A, \delta u^\nu)$ and the principal part given by the $7 \times 7$ matrix
\begin{equation}
\label{princ_linear}
\MA(\Psi,\partial)=\bbm
\left(u^{(0)\alpha}+\xi^{(0)}_{T,\varepsilon} \omega^{(0)\alpha} \right)\partial_\alpha & \xi^{(0)}_{T,{n_V}}\omega^{(0)\alpha}\partial_\alpha & \xi^{(0)}_{T,{n_A}}\omega^{(0)\alpha}\partial_\alpha  & 0_{1\times 4}\\
\left (\frac{1}{3}\Delta^{(0)\mu\alpha}+\xi^{(0)}_{T,\varepsilon} \omega^{(0)\mu} u^{(0)\alpha}\right)\partial_\alpha & \xi^{(0)}_{T,n_V}\omega^{(0)\mu} u^{(0)\alpha}\partial_\alpha & \xi^{(0)}_{T,{n_A}}\omega^{(0)\mu} u^{(0)\alpha}\partial_\alpha  & \frac{1}{2}\xi_T^{(0)} u^{(0)}_{\lambda} \epsilon^{\mu\lambda\alpha}_{\phantom{\mu\lambda\alpha}\nu} u^{(0)\beta}\partial_{\alpha}\partial_\beta \\
\xi^{(0)}_{V,\varepsilon}\omega^{(0)\alpha} \partial_\alpha & \left (u^{(0)\alpha}+\xi^{(0)}_{V,{n_V}}\omega^{(0)\alpha}\right)\partial_\alpha &
\xi^{(0)}_{V,{n_A}}\omega^{(0)\alpha}\partial_\alpha & 0_{1\times 4}\\
\xi^{(0)}_{A,\varepsilon}\omega^{(0)\alpha} \partial_\alpha & 
\xi^{(0)}_{A,{n_V}}\omega^{(0)\alpha}\partial_\alpha & \left (u^{(0)\alpha}+\xi^{(0)}_{A,{n_A}}\omega^{(0)\alpha}\right)\partial_\alpha & 0_{1\times 4}
\ebm,
\end{equation} 
where we introduced the following notation for partial derivatives with respect to thermodynamic fields, e.g., $\xi^{(0)}_{T,\varepsilon}=\partial \xi_T/\partial \varepsilon$ computed at perturbations equal zero. Note that the principal part contains first-order derivatives of $\delta\varepsilon$, $\delta n_V$, and $\delta n_A$ and  second-order derivatives of $\delta u^\nu$. We can now compute the characteristic determinant and find
\begin{eqnarray}
\label{acss}
\det[\MA(\Psi,\varphi)]&=&\det\bbm
b^{(0)}+\xi^{(0)}_{T,\varepsilon} c^{(0)} & \xi^{(0)}_{T,{n_V}}c^{(0)} & {\xi^{(0)}_{T,n_A}}c^{(0)}  & 0_{1\times 4}\\
\frac{1}{3}v^{(0)\mu}+\xi^{(0)}_{T,\varepsilon} b^{(0)}\,\omega^{(0)\mu}  & \xi^{(0)}_{T,n_V}b^{(0)}\,\omega^{(0)\mu}  & \xi^{(0)}_{T,{n_A}}b^{(0)}\,\omega^{(0)\mu}  & \frac{1}{2}\xi^{(0)}_T b^{(0)}\, u^{(0)}_\lambda v^{(0)}_\alpha \epsilon^{\lambda\alpha\mu}_{\phantom{\mu\lambda\alpha}\nu}  \\
\xi^{(0)}_{V,\varepsilon}c^{(0)} & b^{(0)}+\xi^{(0)}_{V,n_V}c^{(0)} &
\xi^{(0)}_{V,n_A}c^{(0)} & 0_{1\times 4}\\
\xi^{(0)}_{A,\varepsilon}c^{(0)} & 
\xi^{(0)}_{A,n_V}c^{(0)} & b^{(0)}+\xi^{(0)}_{A,n_A}c^{(0)} & 0_{1\times 4}
\ebm\nonumber\\
&=&\left (\frac{\xi^{(0)}_T b^{(0)}}{2}\right )^4\det\bbm
b^{(0)}+\xi^{(0)}_{T,\varepsilon} c^{(0)} & \xi^{(0)}_{T,{n_V}}c^{(0)} & {\xi^{(0)}_{T,n_A}}c^{(0)} \\
\xi^{(0)}_{V,\varepsilon}c^{(0)} & b^{(0)}+\xi^{(0)}_{V,n_V}c^{(0)} &
\xi^{(0)}_{V,n_A}c^{(0)} \\
\xi^{(0)}_{A,\varepsilon}c^{(0)} & 
\xi^{(0)}_{A,n_V}c^{(0)} & b^{(0)}+\xi^{(0)}_{A,n_A}c^{(0)}
\ebm \det\left [u^{(0)}_\lambda v^{(0)}_\alpha \epsilon^{\lambda\alpha\mu}_{\phantom{\mu\lambda\alpha}\nu}\right ] =0,
\end{eqnarray}
where we defined
$b^{(0)}=u^{(0)\mu} \varphi_\mu$,  $c^{(0)}=\omega^{(0)\mu} \varphi_\mu$ and $v^{(0)\mu}=\Delta^{(0)\mu\alpha}\varphi_\alpha$. In the last step of  \eqref{acss}, we made use of the fact that the matrix $u^{(0)}_\lambda v^{(0)}_\alpha \epsilon^{\lambda\alpha\mu}_{\phantom{\mu\lambda\alpha}\nu}$ has $u^{(0)}_\nu$ as an eigenvector with zero eigenvalue and, hence, its determinant vanishes.
The vanishing of the characteristic determinant for any $\Sigma = \{ \varphi(x) = 0 \}$ 
implies that the initial-value problem of ideal chiral hydrodynamics in the linear regime is locally ill-posed. As a consequence, it is impossible to find general solutions for these equations of motion. Furthermore, in a relativistic theory, the vanishing of the characteristic
determinant implies acausality. This is because it implies, in particular, that the equations of motion are not hyperbolic and
hyperbolicity is a necessary condition for causality.

\subsection{Nonlinear regime}

For completeness, let us now consider the full nonlinear regime of the system \eqref{csseoms}. One can show that Eqs.~\eqref{csseoms} can be cast in the form \eqref{gen_pde} with the unknowns $\Psi=(\varepsilon,  n_V, n_A, u^\nu)$ with the principal part formally given by the same expression as \eqref{princ_linear}, except that now the background fields are replaced with the dynamical ones. Therefore, the characteristic determinant is given by
\begin{equation}
\det[\MA(\Psi,\varphi)]=\det\bbm
b+\xi_{T,\varepsilon} c & \xi_{T,{n_V}}c & {\xi_{T,n_A}}c  & 0_{1\times 4}\\
\frac{1}{3}v^\mu+\xi_{T,\varepsilon} b\,\omega^\mu  & \xi_{T,n_V}b\,\omega^\mu  & \xi_{T,{n_A}}b\,\omega^\mu  & \frac{1}{2}\xi_T b\, u_\lambda v_\alpha \epsilon^{\lambda\alpha\mu}_{\phantom{\mu\lambda\alpha}\nu}  \\
\xi_{V,\varepsilon}c & b+\xi_{V,n_V}c &
\xi_{V,n_A}c & 0_{1\times 4}\\
\xi_{A,\varepsilon}c & 
\xi_{A,n_V}c & b+\xi_{A,n_A}c & 0_{1\times 4}
\ebm =0,
\end{equation}
where we defined $b=u^{\mu} \varphi_\mu$,  $c=\omega^{\mu} \varphi_\mu$, $v^{\mu}=\Delta^{\mu\alpha}\varphi_\alpha$, defined, e.g., $\xi_{T,\varepsilon}=\partial \xi_T/\partial \varepsilon$, and followed the same steps that we used to prove the result in Eq.~\eqref{acss}. We thus conclude that ideal chiral hydrodynamics defined by the constitutive relations \eqref{css} is ill-posed both in the linear and fully nonlinear regimes. 
Therefore, given that anomalous-free relativistic ideal fluid dynamics has a locally well-posed initial-value problem and causal evolution \cite{ChoquetBruhatGRBook}, one can see that the inclusion of quantum anomaly effects, even in the dissipationless regime, leads to problems that render finding a well-defined general solution of the equations of motion  impossible.

Concerning the physical origin of the problem discussed above, in general, all terms with vorticity in the constitutive relations (including the term proportional to $\omega^\mu u^\nu +\omega^\nu u^\mu$ which makes the system ill-posed) appear from chiral kinetic theory because of the spin-vorticity coupling which, for massless fermions, couples the particle helicity with the fluid vorticity. Therefore, those terms are of quantum origin. In other words, quantum mechanics leads to gradients at the hydrodynamic level even in the absence of dissipation, and it is well-known that gradients may lead to issues when assessing well-posedness, causality, and stability of hydrodynamic theories \cite{Bemfica-Disconzi-Graber-2021,Bemfica:2020zjp}.

\section{Ideal chiral hydrodynamics in the Landau hydrodynamic frame}
\label{secIV}

A given definition of the hydrodynamic fields is called a hydrodynamic frame. There is, of course, an infinite set of hydrodynamic frames \cite{Kovtun:2012rj} with the Landau \cite{LandauLifshitzFluids} and Eckart frames \cite{EckartViscous} being the most well-known definitions. Even in the dissipationless regime, chiral hydrodynamics already contains terms that are of first-order in derivatives and there is  nonzero energy flux in \eqref{css}. We show below that a judicious choice of the hydrodynamic frame is already sufficient to fix the issues displayed by the original formulation of Ref.\ \cite{Chen:2015gta}. We also note that the importance of different hydrodynamic frames in chiral hydrodynamics was already discussed in various contexts, including the definition of Kubo formulas for anomalous transport coefficients~\cite{Landsteiner:2012kd}, and the no-drag frame~\cite{Rajagopal:2015roa,Stephanov:2015roa}.

Let us now  consider \eqref{css} in the so-called Landau frame \cite{LandauLifshitzFluids}. In this case, the flow velocity is defined as an eigenvector of the energy-momentum tensor, a definition that is commonly employed in heavy-ion collision applications \cite{Romatschke:2017ejr}. 
The change to the Landau frame can be done by shifting the velocity \cite{Landsteiner:2012kd,Rajagopal:2015roa,Stephanov:2015roa} as follows $u^\mu = u^\mu_L - \xi_T \omega_L^\mu/(\varepsilon + P)$, where $u^\mu_L$ is the Landau flow velocity and $\omega^\mu_L=\frac12 \epsilon^{\mu\nu\alpha\beta}u_{L\nu} \partial_{\alpha} u_{L\beta}$. Dropping terms of higher order in derivatives, the constitutive relations \eqref{css} become
\begin{subequations}
\label{landau}
\begin{align}
T^{\mu\nu}={}&\varepsilon u_L^\mu u_L^\nu + \Delta_L^{\mu\nu}P , \\
J^\mu_{V}={}& n_V u^\mu_L +\xi_{VL}\omega_L^\mu, \\
J^\mu_{A}={}& n_A u^\mu_L +\xi_{AL}\omega_L^\mu,
\end{align}
\end{subequations}
where $\Delta_L^{\mu\nu}=g^{\mu\nu}+u_L^\mu u_L^\nu$, $\xi_{VL} = \xi_V - \frac{n_V \xi_T}{\varepsilon +P}$ and $\xi_{AL} = \xi_A - \frac{n_A \xi_T}{\varepsilon +P}$. Note that now $u_L^\mu T_{\mu}^\nu = -\varepsilon u_L^\nu$. 
The equations of motion are thus given by
\begin{subequations}
\label{csseomsLandau}
\begin{align}
 u_{L\nu}\partial_\mu T^{\mu\nu}= - D_L \varepsilon - (\varepsilon +P)\theta_L &=0,  \\
 \Delta^{\ \, \alpha}_{L\nu}\partial_\mu T^{\mu\nu}= (\varepsilon + P) D_L u_L^\alpha + \Delta_L^{\alpha\mu}\partial_\mu P  &=0, \\
 \partial_\mu J^\mu_V = D_L n_V +n_V \theta_L + \xi_{VL}\partial_\mu \omega_L^\mu +\omega_L^\mu\partial_\mu \xi_{VL} &=0, \\
 \partial_\mu J^\mu_A = D_L n_A +n_A \theta_L + \xi_{AL}\partial_\mu \omega_L^\mu +\omega_L^\mu\partial_\mu \xi_{AL}  &=0 , 
\end{align}
\end{subequations}
where $D_L=u_L^\mu\partial_\mu$ and $\theta_L=\partial_\mu u^\mu_L$. We will now study the initial-value problem in the linear and nonlinear regimes.

\subsection{Linear regime}

Let us linearize Eqs.~\eqref{csseomsLandau} around a solution of ideal hydrodynamics as done in Sec.~\ref{LinearChiral}. We thus obtain  the following system of equations:
\begin{subequations}
\label{nccLandau}
\begin{align}
\label{eom1_2_lin_ncL}
&D^{(0)} \delta\varepsilon +\delta u_L^\alpha \partial_\alpha \varepsilon^{(0)} + \frac43\varepsilon^{(0)}\partial^\alpha \delta u_{L\alpha} + \frac43 \theta^{(0)} \delta \varepsilon=0 , \\
&\frac43 \varepsilon^{(0)} D^{(0)} \delta u_L^\mu + \frac43\varepsilon^{(0)}\delta u_L^\alpha \partial_\alpha u^{(0)\mu} + \frac43\delta \varepsilon D^{(0)} u^{(0)\mu} + \frac13\Delta^{(0)\mu\alpha}\partial_\alpha \delta \varepsilon + \frac13(u^{(0)\mu} \delta u_L^\alpha+u^{(0)\alpha} \delta u_L^\mu)\partial_\alpha \varepsilon^{(0)} =0,
\label{eom2_2_lin_ncL}\\
&D^{(0)} \delta n_V + \delta u_L^\alpha \partial_\alpha n^{(0)}_V + n^{(0)}_V \partial^\alpha \delta u_\alpha + \delta n_V \theta^{(0)} + \xi_{VL}^{(0)} \partial_\alpha \delta \omega_L^\alpha + \delta\omega_L^\alpha\partial_\alpha \xi_{VL}^{(0)} + \omega^{(0)\alpha}\partial_\alpha \delta\xi_{VL} + \delta\xi_{VL} \partial_\alpha \omega^{(0)\alpha} =0,
\label{eom3_2_lin_ncL} \\
&D^{(0)} \delta n_A + \delta u_L^\alpha \partial_\alpha n^{(0)}_A + n^{(0)}_A \partial^\alpha \delta u_\alpha + \delta n_A \theta^{(0)} + \xi_{AL}^{(0)} \partial_\alpha \delta \omega_L^\alpha + \delta\omega_L^\alpha\partial_\alpha \xi_{AL}^{(0)} + \omega^{(0)\alpha}\partial_\alpha \delta\xi_{AL} + \delta\xi_{AL} \partial_\alpha \omega^{(0)\alpha} =0. \label{eom4_2_lin_ncL}
\end{align}
\end{subequations}
Equations \eqref{nccLandau} form a system of first-order PDEs with unknowns $\Psi=(\delta\varepsilon, \delta n_V, \delta n_A, \delta u_L^\nu)$.
For simplicity, let us first consider the case with vanishing vector chemical potential, which implies $J^\mu_V=0$ \cite{Chen:2015gta,Yang:2018lew}, but nonzero axial chemical potential $\mu_A$. In this case the principal part reads 
\begin{equation}
\label{princLandau_one}
\mathcal{A}(\Psi, \partial)=\begin{bmatrix}
 u^{(0)\alpha} & 0 & \frac43 \varepsilon^{(0)} \delta^\alpha_\nu \\
\frac13 \Delta^{(0)\mu\alpha} & 0 & \frac43 \varepsilon^{(0)} u^{(0)\alpha} \delta^{\mu}_\nu \\
\xi^{(0)}_{AL,\varepsilon} \omega^{(0)\alpha} & (u^{(0)\alpha} +\xi^{(0)}_{AL,n_A} \omega^{(0)\alpha}) & {H}^{(0)\alpha}_{\ \ \ \, \nu}  
\end{bmatrix}\partial_\alpha,
\end{equation}
where
\begin{equation}
{H}^{(0)\alpha}_{\ \ \ \, \nu} = n^{(0)}_{A} \delta^\alpha_\nu + \frac12 (\xi^{(0)}_{AL}\partial_\rho u^{(0)}_{L\tau}+u^{(0)}_{L\tau}  {\partial}_{\rho} \xi^{(0)}_{AL}) \epsilon^{\rho\tau\alpha}_{\ \ \ \ \nu} .
\end{equation}
The characteristic determinant reads
\beq
\label{eq:detlandaulin}
\det[ \MA(\Psi,\varphi)]=\left(\frac{4}{3}\varepsilon^{(0)}\right)^4 (b^{(0)})^3\left [(b^{(0)})^2- \frac13 (v^{(0)})^2\right ](b^{(0)}+\xi^{(0)}_{AL,n_A}c^{(0)}),
\eeq
where $b^{(0)}=u^{(0)}_\mu \varphi^\mu$, $v^{(0)}_\mu =\Delta^{(0)}_{\mu\nu}\varphi^\nu$, $c^{(0)}=\omega^{(0)}_\mu \varphi^\mu$ and  $\varepsilon^{(0)}>0$.
We notice that the factors in \eqref{eq:detlandaulin} have the same structures as the examples discussed in Sec.~\ref{sec:initial}. Specifically $(b^{(0)})^2-\frac13 (v^{(0)})^2=0$ has the same form as the characteristic determinant of the wave equation \eqref{wave} which leads to causal roots,  $b^{(0)}=0$ is causal. The remaining root comes from
\begin{equation}
\label{adv_omega}
b^{(0)}+\xi^{(0)}_{AL,n_A} c^{(0)}=\varphi^\mu(u^{(0)}_\mu+\xi^{(0)}_{AL,n_A} \omega^{(0)}_\mu)=0 
\end{equation}
which has the same form as the characteristic determinant of the advection equation. Since $\xi^{(0)}_{AL,n_A}$ is real, we have that (i) in \eqref{condvarphi} automatically holds. Thus, we can conclude that Eq.~\eqref{adv_omega} leads to causal roots if, and only if, the vector $(u_0^\mu+\xi^{(0)}_{AL,n_A} \omega_0^\mu)$ is timelike or lightlike, namely
\begin{equation}
\label{conditionsm1}
|\xi^{(0)}_{AL,n_A}|\, \omega^{(0)}_L\le 1,
\end{equation}
where $\omega^{(0)}_L=\sqrt{\omega_L^{(0)\mu}\omega^{(0)}_{L\mu}}$ (recall that $\omega^{(0)\mu}_L$ is spacelike).
Using the results in \cite{DisconziFollowupBemficaNoronha}, one obtains local well-posedness in Gevrey spaces. Thus Eq.~\eqref{conditionsm1} is the condition for causality and local well-posedness. We note that Eq.~\eqref{conditionsm1} sets a bound for the vorticity strength in terms of the anomaly coefficient\footnote{Using the results in \cite{Chen:2015gta,Yang:2018lew}, one can show that Eq.\ \eqref{conditionsm1} becomes simply $6\omega^{(0)}_L[5(\mu_A^{(0)})^3 + \pi^2 \mu_A^{(0)} (T^{(0)})^2]/[15(\mu^{(0)}_A)^4+6\pi^2 (\mu^{(0)}_A)^2 (T^{(0)})^2 + 7\pi^4 (T^{(0)})^4]\leq 1$.}. 
This provides a clear example in which quantum effects restrict how fast fluids can spin in relativity.

For completeness, let us now consider the general case where $J^\mu_V$ is also present. Using similar steps as before, we obtain
\beq
\label{det0000}
\det[ \MA(\Psi,\varphi)]=\left(\frac{4}{3}\varepsilon^{(0)}\right)^4 (b^{(0)})^{3}\left ((b^{(0)})^2- \frac13 (v^{(0)})^2\right )(b^{(0)}-\alpha^{(0)}_+ c^{(0)})(b^{(0)}-\alpha^{(0)}_- c^{(0)}),
\eeq
where
\beq
\alpha^{(0)}_\pm=\frac{-(\xi^{(0)}_{AL,n_A}+\xi^{(0)}_{VL,n_V})\pm\sqrt{(\xi^{(0)}_{AL,n_A}-\xi^{(0)}_{VL,n_V})^2+4\xi^{(0)}_{AL,n_V}\xi^{(0)}_{VL,n_A}}}{2}.
\eeq 
The conditions read
\begin{subequations}
\label{conditions}
\bea
&&(\xi^{(0)}_{AL,n_A}-\xi^{(0)}_{VL,n_V})^2+4\xi^{(0)}_{VL,n_V}\xi^{(0)}_{VL,n_A}\ge0,\label{cond2}\\
&&\left [-\left(\xi^{(0)}_{AL,n_A}+\xi^{(0)}_{VL,n_V}\right )+\sqrt{(\xi^{(0)}_{AL,n_A}-\xi^{(0)}_{VL,n_V})^2+4\xi^{(0)}_{AL,n_V}\xi^{(0)}_{VL,n_A}}\right ]\omega^{(0)}_L\le 2,\label{cond3}\\
&&\left [-\left (\xi^{(0)}_{AL,n_A}+\xi^{(0)}_{VL,n_V}\right )-\sqrt{(\xi^{(0)}_{AL,n_A}-\xi^{(0)}_{VL,n_V})^2+4\xi^{(0)}_{AL,n_V}\xi^{(0)}_{VL,n_A}}\right ]\omega^{(0)}_L\ge -2.\label{cond4}
\eea
\end{subequations}
Concerning (i) in \eqref{condvarphi}, the roots are real if, and only if, $\alpha^{(0)}_\pm$ are real, i.e., if \eqref{cond2} is verified. Furthermore, in analogy with what has been done above, the condition to guarantee (ii) and that the determinant~\eqref{det0000} is different from zero is that the vectors are timelike or lightlike, namely $|\alpha^{(0)}_\pm|\omega^{(0)}_L\le1$ which corresponds  to \eqref{cond3} and \eqref{cond4}. Again, one obtains local well-posedness in Gevrey spaces. Thus, the Landau frame formulation in the linear regime \eqref{nccLandau} admits a causal and well-posed initial-value problem.

\subsection{Nonlinear regime}

\label{sec:landau-nl}

The equations of motion \eqref{csseomsLandau} are first-order PDEs, but they are fully nonlinear rather than quasilinear since there are nonlinear first-order derivative terms (i.e., $\sim (\partial \Psi)^2$, as the ones coming from the conservation of the currents $J^\mu_{V/A}$ which are of the form $\partial_\mu \omega_L^\mu=\frac{1}{2}\epsilon^{\mu\nu\alpha\beta}\partial_\mu u_{L\nu}\partial_\alpha u_{L\beta}$ and $\omega^\mu \partial_\mu \xi_{VL/AL}$). This implies that local well-posedness and causality cannot be studied in a straightforward way using  standard techniques which apply to quasilinear systems as discussed above. However, a more detailed analysis reveals that \eqref{landau} is causal 
and locally well-posed in Gevrey spaces if the same conditions as those in the linear case hold, namely Eqs.~\eqref{conditions}, where the background quantities are now replaced by the full fields (e.g., $\omega^{(0)}_L$ is replaced with $\omega_L$). Below we give the proof of such statement.

If we assume that the initial data are prescribed along a hypersurface at constant given initial time $x^0=t_0$, the extension to a generic initial condition is straightforward. Hence, the initial data are given by $\Psi(t_0,x^i)$. In order to prove well-posedness and causality we
proceed in two steps. The first step consists of finding the conditions for expressing $\partial_0\Psi(t_0,x^i)$ in terms of $\Psi(t_0,x^i)$ and $\partial_i\Psi(t_0,x^i)$. This is a minimal requirement for a solution to be found. Since we want to study the evolution given initial data prescribed on a general hypersurface (not necessarily at constant time), it is convenient to work in a covariant way. We decompose the derivative into a parallel and perpendicular part with respect to a normalized time-like vector $q^\mu$ ($q^\mu q_\mu =-1$), i.e.,
\begin{equation}
\label{partialq}
\partial^\alpha = -q^\alpha \tilde{D} + \tilde{\nabla}^\alpha,
\end{equation}
with $\tilde{D}=q^\alpha \partial_\alpha$ and $\tilde{\nabla}^\mu=(g^{\mu\nu}+q^\mu q^\nu)\partial_\nu$. Hence, our system can be written as
\begin{equation}
-M^\alpha q_\alpha \tilde{D} \Psi + F =0,
\end{equation}
where $\Psi=(\varepsilon,n_V,n_A,u_L^\nu)$ and $F$ are $7\times 1$ column matrices, with $F$ containing all the remaining terms that does not contain a time derivative along $q^\alpha$ of the fields $\Psi$, and $M^\alpha$ are $7\times 7$ matrices. Thus, we search for the conditions such that the matrix $-q_\alpha M^\alpha$ is invertible, namely $\det (-q_\alpha M^\alpha)\neq 0$. For the sake of simplicity, let us first consider the case of vanishing vector chemical potential which implies $n_V=0$ and $J^\mu_V=0$. The matrix $-q_\alpha M^\alpha$ is now $6\times 6$ and it reads
\begin{equation}
\label{35}
-q_\alpha M^{\alpha}=\begin{bmatrix}
 \tilde{b} & 0 & \frac43 \varepsilon q_{\nu} \\
\frac13 v^\mu & 0 & \frac43 \varepsilon \tilde{b} \delta^{\mu}_\nu \\
\xi_{AL,\epsilon} \tilde{c} & (\tilde{b} +\xi_{AL,n_A} \tilde{c}) & \tilde{H}_\nu  
\end{bmatrix},
\end{equation}
where we made used of the conformal equation of state $P=\varepsilon/3$ and defined  $\tilde{b}=u_{L\mu} q^\mu$, $\tilde{v}^\mu = \Delta^{\mu\alpha}q_\alpha$, $\tilde{c}=\frac12 q_\mu\epsilon^{\mu\nu\alpha\beta}u_{L\nu} (\tilde{\nabla}_\alpha u_{L\beta})$, and 
\begin{equation}
\tilde{H}^\nu = n_{A} q^\nu + \frac12 (2\xi_{AL}\tilde{\nabla}_\alpha u_{L\beta}+u_{L\beta}  \tilde{\nabla}_{\alpha} \xi_{AL}) q_\mu\epsilon^{\alpha\beta\mu\nu}.
\end{equation}
The determinant of the matrix in \eqref{35} is given by
\begin{align}
\label{det1}
\det(-q_\alpha M^{\alpha})= \left(\frac43 \varepsilon \right)^4 \tilde{b}^3  \left( \tilde{b}^2 - \frac13\tilde{v}^2 \right)(\tilde{b} +\xi_{AL,n_A} \tilde{c}),  
\end{align}
which is different from zero if
\begin{equation}
\label{Cond11}
\tilde{b} +\xi_{AL,n_A} \tilde{c} = q_{\mu}(u_L^\mu+\xi_{AL,n_A} \omega_L^\mu) \neq 0
\end{equation}
for any timelike $q^\mu$.

The second step of the proof consists of converting the first-order system \eqref{csseomsLandau} into a quasilinear second-order one, so that standard techniques to study well-posedness and causality can be applied \cite{Courant_and_Hilbert_book_2}. This can be achieved by acting with  $u^\alpha\partial_\alpha$ onto \eqref{csseomsLandau}. In this way, we obtain the new second-order system which is of the form \eqref{gen_pde}, where the $6\times 6$ matrix linear differential operator of the principal part is given by
\beq
\label{AA}
\MA(\Psi,\partial)=\bbm
u_L^\alpha u_L^\beta & 0 & \frac{4}{3}\varepsilon u_L^{(\alpha}\delta^{\beta)}_\nu \\
\frac13 u_L^{(\alpha}\Delta^{\mu\beta)} & 0_{4\times 1} & \frac{4}{3}\varepsilon u_L^\alpha u_L^\beta\delta^\mu_\nu\\
\xi_{AL,\varepsilon}  u_L^{(\alpha}\omega_L^{\beta)} & u_L^\alpha u_L^\beta+\xi_{AL,n_A} u_L^{(\alpha}\omega_L^{\beta)} & H^{\alpha\beta}_\nu
\ebm \partial_\alpha\partial_\beta,
\eeq
with the notation, e.g., $u^{(\alpha}_L\omega_L^{\beta)}=(1/2)(u_L^\alpha \omega_L^\beta + u_L^\beta \omega_L^\alpha)$, and 
\begin{equation}
H^{\alpha\beta}_\nu = n_A u_L^{(\alpha}\delta^{\beta )}_\nu + \frac12 (2\xi_{AL}\partial_\mu u_\lambda + u_\lambda \partial_\mu \xi_{AL})\epsilon^{\mu\lambda(\alpha}_{\ \ \ \ \ \nu} u^{\beta )}.    
\end{equation}
The characteristic determinant of \eqref{AA} is given by
\beq
\det[ \MA(\Psi,\varphi)]=\left(\frac{4}{3}\varepsilon\right)^4 b^{9}\left (b^2- \frac13 v^2\right )(b+\xi_{AL,n_A}c),
\eeq
where $b$, $v^\mu$, and $c$ are defined as in the main text with $u^\mu_L$ instead of $u^\mu$.
This determinant has the same structure as \eqref{det0000} and we can thus conclude that the roots $b=0$ and $b^2-\frac13 v^2=0$ are causal. The remaining root comes from the equation
\begin{equation}
\label{Cond22}
   b+\xi_{AL,n_A} c=\varphi_{\mu}(u_L^\mu+\xi_{AL,n_A} \omega_L^\mu)=0 .
\end{equation}
In order for conditions \eqref{Cond11} and \eqref{Cond22} to be satisfied for any timelike $q_\mu$ and for any $\varphi_\mu$ such that (ii) in \eqref{condvarphi} is obeyed, the vector $(u_L^\mu+\xi_{AL,n_A} \omega_L^\mu)$ must be timelike or lightlike, namely Eq.~\eqref{conditionsm1} in the main text must hold. Using the results in \cite{DisconziFollowupBemficaNoronha}, one obtains local well-posedness in Gevrey spaces.

Let us now consider the general case where $J^\mu_V$ is also present. Using similar steps as before, we obtain
\begin{align}
\label{det1000}
\det(-q_\alpha M^{\alpha})= \left(\frac43 \varepsilon \right)^4 \tilde{b}^3  \left( \tilde{b}^2 - \frac13\tilde{v}^2 \right)(\tilde{b}-\alpha_+ \tilde{c})(\tilde{b}-\alpha_- \tilde{c}),  
\end{align}
where
\beq
\alpha_\pm=\frac{-(\xi_{AL,n_A}+\xi_{VL,n_V})\pm\sqrt{(\xi_{AL,n_A}-\xi_{VL,n_V})^2+4\xi_{AL,n_V}\xi_{VL,n_A}}}{2}.
\eeq 
After applying the operator $u_\alpha \partial^\alpha$, the determinant of the principal part is now given by
\beq
\det[ \MA(\Psi,\varphi)]=\left(\frac{4}{3}\varepsilon\right)^4 b^{10}\left (b^2- \frac13 v^2\right )(b-\alpha_+ c)(b-\alpha_- c).
\eeq
Concerning (i), the roots are real if, and only if, $\alpha_\pm$ are real, i.e., if \eqref{cond2} is verified. Furthermore, in analogy with what has been done above, the condition to guarantee (ii) and that the determinant~\eqref{det1000} is different from zero is that the vectors are timelike, namely $|\alpha_\pm|\omega_L\le1$ which corresponds  to \eqref{cond3} and \eqref{cond4}. Again, one obtains local well-posedness in Gevrey spaces.

We thus see that the Landau frame formulation of ideal chiral kinetic theory admits a well-posed initial-value problem both at the linear and full nonlinear regimes, with causal evolution, as long as the nontrivial conditions \eqref{conditions}  that effectively place a bound on the vorticity are satisfied. Therefore, ideal chiral hydrodynamics defined by \eqref{landau} can be solved and numerical simulations of such a fluid can be performed (one can test whether inequalities \eqref{conditions} hold at each time step in order
to verify whether causality holds). Finally, our results show that quantum effects in chiral fluids influence the choice of the hydrodynamic frame, even in the absence of dissipation.

\section{Acausality of viscous chiral hydrodynamics} 
\label{secV}

For completeness, we now investigate the case of viscous chiral hydrodynamics following  the entropy-current analysis of Ref.\ \cite{Son:2009tf}, and write the energy-momentum tensor and the current in this way
\begin{subequations}
\label{ss}
\begin{align}
T^{\mu\nu}&= \varepsilon u^\mu u^\nu + (P - \zeta \partial_\lambda u^\lambda)\Delta^{\mu\nu} -2 \eta \sigma^{\mu\nu} , \\
J^\mu&= n u^\mu -\sigma T \Delta^{\mu\nu}\partial_\nu \left(\frac \mu T\right) + \sigma E^\mu + \xi\omega^\mu + \xi_B B^\mu,
\end{align}
\end{subequations}
where $n$ is the particle number density and the shear tensor is $\sigma^{\mu\nu}=\Delta^{\mu\nu\alpha\beta} \partial_\alpha u_\beta$ with $\Delta^{\mu\nu\alpha\beta}=(\Delta^{\mu\alpha}\Delta^{\nu\beta}+\Delta^{\mu\beta}\Delta^{\nu\alpha})/2-\Delta^{\mu\nu}\Delta^{\alpha\beta}/3$. Also, we defined the covariant electric and magnetic fields, $E^\mu = F^{\mu\nu}u_\nu$ and $B^\mu= (1/2)\epsilon^{\mu\nu\alpha\beta} u_\nu F_{\alpha\beta}$, respectively, with $F^{\mu\nu}$ being the electromagnetic field tensor (assumed to be nondynamical \cite{Son:2009tf}). The coefficients $\zeta$ and $\eta$ are the bulk and shear viscosities, respectively, while $\sigma$ is the conductivity. Following  \cite{Son:2009tf}, we restrict ourselves here to the case of a single axial current with the associated chemical potential $\mu$ (the generalization involving multiple currents, as in the previous section, is known \cite{Sadofyev:2010pr,Neiman:2010zi}). 

We note that the hydrodynamic fields in \eqref{ss} are expressed in the Landau hydrodynamic frame \cite{LandauLifshitzFluids}. In the presence of the anomaly, the conservation laws are given by
\begin{equation}
\label{hydro_anomaly}
\partial_\mu T^{\mu\nu} = F^{\nu\lambda}J_\lambda, \qquad \partial_\mu J^{\mu}= C E_\mu B^\mu ,
\end{equation}
where $C$ is the anomaly coefficient, which determines the coefficients $\xi$ and $\xi_B$ (see \cite{Son:2009tf}). We prove below that the set of nonlinear PDEs given by \eqref{hydro_anomaly} violates causality.  

We again notice that the set of second-order PDEs in \eqref{hydro_anomaly} is a quasilinear system of the form  \eqref{gen_pde}. We use the projections of the first equation in  \eqref{hydro_anomaly}, i.e., $u_\nu( \partial_\mu T^{\mu\nu} - F^{\nu\mu} J_\mu) = 0$ and $\Delta_\nu^\alpha( \partial_\mu T^{\mu\nu} -  F^{\nu\mu}J_\mu)=0$. Also, we choose the unknowns to be $\Psi=(\varepsilon, \mu/T, u^\nu)$. The principal part is given by
\begin{equation}
\MA(\Psi,\partial)=\bbm
u^\alpha\partial_\alpha & 0 & 0_{1\times 4}\\
C^{\mu\alpha} \partial_\alpha & 0_{4\times 1} & -D^{\mu\alpha\beta}_\nu \partial_\alpha \partial_\beta\\
G^\alpha\partial_\alpha & -\sigma T\Delta^{\alpha\beta}\partial_\alpha \partial_\beta & 0
\ebm ,
\end{equation}
where
\bea
C^{\mu\alpha}&=&(P_\varepsilon-\zeta_\varepsilon\partial_\nu u^\nu)\Delta^{\mu\alpha}-2\eta_\varepsilon\sigma^{\mu\alpha},\\
D^{\mu\alpha\beta}_\nu&=&\left (\zeta+\frac{\eta}{3}\right )\Delta^{\mu(\alpha}\delta^{\beta)}_\nu+\eta\Delta^{\alpha\beta}\delta^\mu_\nu,\\
G^\alpha&=&n_\varepsilon u^\alpha-(\sigma T)_\varepsilon\Delta^{\alpha\beta}\partial_\beta\left(\frac \mu T\right)+\sigma_\varepsilon E^\alpha+\xi_\varepsilon+\xi_{B,\varepsilon} B^\alpha
\eea
and we made use again of the notation for the partial derivatives, e.g., $P_\varepsilon=\partial P/\partial\varepsilon$. One can show that the characteristic determinant is given by
\bea
\det[\MA(\Psi,\varphi)]
&=&-\sigma T (\Delta_{\mu\nu}\varphi^\mu \varphi^\nu)^5 b\left (\zeta+\frac{4\eta}{3}\right ).
\label{sschar}
\eea
This result implies acausality, since $v^2=\Delta^{\alpha\beta}\varphi_\alpha\varphi_\beta=0$ corresponds to $\varphi_\alpha\varphi^\alpha=-b^2+v^2=-b^2<0$ which, in turn, implies that the system is not hyperbolic \cite{ChoquetBruhatGRBook}.  Therefore, the constitutive relations defined by \eqref{ss} do not provide a viable causal viscous generalization of ideal chiral hydrodynamics. In particular, its inherent acausal nature forbids numerical simulations of such theory in relativistic fluids, such as the quark-gluon plasma.

We observe that standard Landau-Lifshitz theory \cite{LandauLifshitzFluids} corresponds to the limit of the constitutive relations \eqref{ss} in which both $\xi$ and $\xi_B$ vanish (with $F^{\mu\nu}=0$). In this regard, it is well-known that Landau-Lifshitz theory displays acausal behavior when linearized around equilibrium \cite{Hiscock_Lindblom_instability_1985}. However, one can show that Landau-Lifshitz theory has the same characteristic determinant as in \eqref{sschar}, which implies that this theory is acausal also in the nonlinear regime. Finally, we note that \eqref{ss} and Landau-Lifshitz theory become identical when linearized around equilibrium (for $F^{\mu\nu}=0$) so \eqref{ss} also suffers from the same unphysical instabilities known to plague Landau-Lifshitz theory \cite{Hiscock_Lindblom_instability_1985}, which render such formulation unsuited for numerical simulations.      


Even though \cite{Bemfica:2017wps,Kovtun:2019hdm,Bemfica:2019knx,Hoult:2020eho,Bemfica:2020zjp} initially considered only anomaly-free theories, the same arguments used there should apply when investigating the macroscopic evolution of relativistic fluids in the presence of quantum anomalies. All the possible terms of first-order in spacetime derivatives, which in this case will  involve also $\omega_\mu$, must be included when writing the most general constitutive relations that define the energy-momentum tensor and the currents. Since the simple first-order shift to the Landau frame was shown here to cure the issues displayed by ideal chiral hydrodynamics \eqref{css}, one can see that  the choice of the hydrodynamic frame should play a key role in chiral hydrodynamics. Therefore, in the presence of quantum anomalies, questions concerning the nature of hydrodynamic frames are unavoidable and crucial already in the dissipationless regime.

The inclusion of all the possible first-order terms should naturally fix the acausal nature displayed by chiral viscous hydrodynamics \eqref{ss}, as it did in the anomalous-free case. In particular, the presence of terms such as $u^\alpha \partial_\alpha T$ and $u^\alpha \partial_\alpha \mu$ in the constitutive relations fundamentally changes the structure of the equations, turning the system hyperbolic and well-posed. A detailed study of causality, stability, and well-posedness in this case is extremely lengthy and complex (see \cite{Bemfica-Disconzi-Graber-2021,Bemfica:2020zjp} for a simpler example), and it will be presented elsewhere.

\section{Conclusions} 
\label{conclusions}

In this paper we investigated causality and the initial-value problem of the ideal chiral hydrodynamic equations derived from kinetic theory \cite{Chen:2015gta}. We performed a comprehensive study of such properties both in the linear and nonlinear regimes of these equations. The linear regime describes perturbations around a general (e.g. nonlinear) solution of the ideal hydrodynamic equations, which includes the rotating global equilibrium state. We found that ideal chiral hydrodynamics, in a general hydrodynamic frame where energy diffusion is nonzero,  has an ill-posed initial-value problem and is acausal. This issue appears both in the linear and nonlinear regimes. Having an ill-posed initial-value problem means that a general solution of the partial differential equations does not exist or is not unique, which implies that the system cannot be solved and numerical simulations cannot be performed.
However, when using other choices of hydrodynamic frames, namely the Landau frame, the theory becomes locally well-posed 
and causal if certain conditions are met. Thus, the version in Eq.\ \eqref{landau} of the theory proposed by \cite{Chen:2015gta} is free from unphysical features and can be numerically solved. In particular, we found that the magnitude of the vorticity is directly constrained by the coefficient that encodes the anomaly due to causality. This provides a concrete example where the size of the gradients of classical, macroscopic quantities (i.e., the vorticity) is bound by the relativistic quantum nature of the fluid constituents. These constraints appear both in the linear and nonlinear regimes.

Furthermore, for completeness, we also studied the well-known first-order formulation of viscous chiral hydrodynamics in the Landau frame constrained from the entropy-current  \cite{Son:2009tf}, and proved that this theory is also acausal, which should be expected given that the nonanomalous terms in the equations reduce to Navier-Stokes theory, which is known to be acausal and unstable. We argued that such issues can be naturally fixed in the derivative expansion by taking into account all the possible terms at first-order, in the context of \cite{Bemfica:2017wps,Kovtun:2019hdm,Bemfica:2019knx,Hoult:2020eho,Bemfica:2020zjp}.


Our work shows that the choice of hydrodynamic frame plays an important role in chiral hydrodynamics already in the dissipationless limit due to quantum effects. Indeed, a bad choice of hydrodynamic frame can render the theory not only acausal but also ill-posed. Additionally, our results illustrate that defining local thermodynamic equilibrium in relativistic fluids where chirality and vorticity are considered is  nontrivial and deserves further investigation. We note that conceptual problems related to the definition of local equilibrium for hydrodynamic theories where spin degrees of freedom are promoted to dynamical variables (the so-called spin hydrodynamics) have been pointed out in different works before \cite{Montenegro:2017rbu,Montenegro:2018bcf,Florkowski:2017ruc,Hattori:2019lfp,Weickgenannt:2020aaf} (see also related work \cite{Montenegro:2017lvf,Becattini:2018duy,Florkowski:2018fap,Bhadury:2020puc,Speranza:2020ilk,Shi:2020htn,Fukushima:2020ucl,Li:2020eon,Gallegos:2021bzp}).    

Finally, it would be interesting to investigate causality and the initial-value formulation in approaches to chiral hydrodynamics that differ from the derivative expansion, such as Israel-Stewart theory \cite{MIS-6}. Reference \cite{Gorbar:2017toh} investigated chiral hydrodynamics in this approach, but their analysis was mostly limited to the linearized  regime around a rotating global equilibrium state. Using the results from Ref.\ \cite{Bemfica:2020xym}, a general analysis of the nonlinear behavior of Israel-Stewart-like chiral hydrodynamic theories (written in a general hydrodynamic frame \cite{Noronha:2021syv}) is possible. This challenging problem is left for future work.

\section*{Acknowledgments} 
The authors thank A. Sadofyev for useful discussions, and W. Zajc for valuable comments about the constraints on the vorticity strength.
MMD is partially supported by a Sloan Research Fellowship provided by the Alfred P. Sloan Foundation, by NSF grant DMS-2107701, and by a Dean's Faculty Fellowship. 
JN is partially supported by the U.S. Department of Energy, Office of Science, Office for Nuclear Physics under Award No. DE-SC0021301.

\appendix

\section{Equations of motion}
\label{appendixA}

The  system \eqref{csseoms} in the linearized case discussed in Sec. \ref{LinearChiral} takes the form
\begin{subequations}
\label{ncc}
\begin{align}
\label{eom1_2_lin_nc}
&u^{(0)\alpha} \partial_\alpha \delta\epsilon +\delta u^\alpha \partial_\alpha \epsilon^{(0)} + \frac43\epsilon^{(0)}\partial^\alpha \delta u_\alpha + \frac43 \delta \epsilon \partial^\alpha u^{(0)}_{\alpha} + \partial^\alpha (\xi^{(0)}_T\omega^{(0)}_\alpha)+ \partial^\alpha (\delta\xi_T\omega^{(0)}_\alpha) + \partial^\alpha (\xi^{(0)}_T\delta\omega_\alpha) + \xi_T^{(0)}\omega^{{(0)}\alpha}u_\beta^{(0)}\partial^\beta u^{(0)}_\alpha\nonumber\\
& + \delta\xi_T\omega^{{(0)}\alpha}u_\beta^{(0)}\partial^\beta u^{(0)}_\alpha + \xi^{(0)}_T\delta\omega^{\alpha}u_\beta^{(0)}\partial^\beta u^{(0)}_\alpha + \xi_T^{(0)}\omega^{{(0)}\alpha}\delta u_\beta\partial^\beta u^{(0)}_\alpha =0, \\
&\frac43 \epsilon^{(0)} D^{(0)} \delta u^\mu + \frac43\epsilon^{(0)}\delta u^\alpha \partial_\alpha u^{(0)\mu} + \frac43\delta \epsilon D^{(0)} u^{(0)\mu} + \frac13\Delta^{(0)\mu\alpha}\partial_\alpha \delta \varepsilon + \frac13(u^{(0)\mu} \delta u^\alpha+u^{(0)\alpha} \delta u^\mu)\partial_\alpha \varepsilon^{(0)} \nonumber \\
& + \xi_T^{(0)}\omega^{(0)\alpha}\partial_\alpha u^{(0)\mu} + \delta\xi_T\omega^{(0)\alpha}\partial_\alpha u^{(0)\mu} + \xi_T^{(0)}\delta\omega^{\alpha}\partial_\alpha u^{(0)\mu} + \xi_T^{(0)}\omega^{(0)\alpha}\partial_\alpha \delta u^{\mu} + \Delta^{(0)\mu}_{\ \ \ \alpha} D^{(0)}(\xi_T^{(0)}\omega^{(0)\alpha}) +\delta\Delta^{\mu}_{\alpha} D^{(0)}(\xi_T^{(0)}\omega^{(0)\alpha})  \nonumber\\
& + \Delta^{(0)\mu}_{\ \ \ \alpha} \delta D(\xi_T^{(0)}\omega^{(0)\alpha})+ \Delta^{(0)\mu}_{\ \ \ \alpha} D^{(0)}(\delta\xi_T\omega^{(0)\alpha}) + \Delta^{(0)\mu}_{\ \ \ \alpha} D^{(0)}(\xi_T^{(0)}\delta\omega^{\alpha}) + \xi_T^{(0)}\omega^{(0)\mu}\theta^{(0)} + \delta\xi_T\omega^{(0)\mu}\theta^{(0)} \nonumber \\
&+ \xi_T^{(0)}\delta\omega^{\mu}\theta^{(0)}+ \xi_T^{(0)}\omega^{(0)\mu}\delta\theta =0 ,
\label{eom2_2_lin_nc} \\
&u^{(0)\mu} \partial_\mu \delta n_V + \delta u^\mu \partial_\mu n^{(0)}_V + n^{(0)}_V \partial^\mu \delta u_\mu + \delta n_V \partial_\mu u^{(0)\mu} + \xi_V^{(0)} \partial_\mu \delta \omega^\mu + (\partial_\mu \xi_V^{(0)})\delta\omega^\mu + (\partial_\mu \delta\xi_V)\omega^{(0)\mu} + \delta\xi_V \partial_\mu \omega^{(0)\mu} =0,
\label{eom3_2_lin_nc} \\
&u^{(0)\mu} \partial_\mu \delta n_A + \delta u^\mu \partial_\mu n^{(0)}_A + n^{(0)}_A \partial^\mu \delta u_\mu + \delta n_A \partial_\mu u^{(0)\mu} + \xi_A^{(0)} \partial_\mu \delta \omega^\mu + (\partial_\mu \xi_A^{(0)})\delta\omega^\mu + (\partial_\mu \delta\xi_A)\omega^{(0)\mu} + \delta\xi_A \partial_\mu \omega^{(0)\mu} =0. \label{eom4_2_lin_nc}
\end{align}
\end{subequations}

\bibliography{References_chiral.bib}{}

\end{document}